\documentclass[
    ,final            
  ]
  {aipproc}

\layoutstyle{6x9}

\begin{document}

\title{\hskip10cm NT@UW-10-07\\Travels With Tony-Nucleon Structure Through Our Ages}

\classification{
                \texttt{13.40.Gp, 13.60.Hb, 14.20.Dh, 24.85.+p }}

\author{Gerald A. Miller}{
  address={Physics Department, University of Washington, Seattle, WA, USA 98195-1560}
}


\begin{abstract}
Nucleon structure is currently understood from a unified light-front, infinite-momentum-frame framework. 
The specific examples of the neutron transverse charge distribution and the shape of the proton are 
discussed here.
\end{abstract}

\maketitle
\newcommand{\bfgamma}{\mbox{\boldmath $\gamma$}}
\newcommand{\bfsigma}{\mbox{\boldmath $\sigma$}}

The theme of this talk is that   ways of  examining  nucleon structure have changed with time.
The period of 1979-1982 was an exciting time \cite{cbm1,cbm2,cbm3} and so is now.  In this talk, I'll discuss how 
the neutron charge distribution and  the shape of the proton are determined through transverse densities and transverse momentum distributions. 

Before going into the physics, I'd like to convey the most important message: Happy Birthday Tony!

Here is an outline of this talk. I'll begin by discussing the   cloudy bag model briefly, Then I'll discuss some
phenomenology that shows that the  proton is   not round.  While this result is based on a model, some new model-independent techniques can be brought to bear on many aspects of nucleon structure. Therefore I'll discuss   
the model independent   neutron charge transverse density \cite{myneutron}. Then I'll return to the shape of the proton\cite{shape1,shape2,transverse}.  This can be
measured using lattice QCD through a type of  impact parameter dependent GPD, and in    experiment through a Transverse Momentum Distribution, TMD. The  impact parameter dependent GPD  is a coordinate-space density while the TMD is a momentum space density. All of the subjects in this talk are discussed in more detailed fashion in the recent review
  \cite{transverse}.
\section{Cloudy Bag Model}
I haven't written about the Cloudy Bag Model CBM in a long time, but this workshop is a special occasion. In the CBM, the nucleon
is treated as a confined system of quarks (modeled by the MIT bag) that interact with a cloud of pions. The use of this model led to 
many successful predictions \cite{tonyrev,millerrev}. One feature of the model is that the  pion penetrates to the bag interior.

A special  example involves the neutron. Sometimes the neutron makes a quantum fluctuation into a proton and $\pi^-$.
The cloudy bag model calculation of Fourier transform  of the neutron electric form factor $=4\pi \int Q^2dQ G_{En}(Q^2) \sin(Qr)/(Qr)$ is shown
in Fig. 11 of \cite{cbm3}. The figure shows a positive contribution at $r=0$ and a negative contribution  for values of $r$ corresponding to positions outside of the bag. In recent years, I have written that the beautiful picture and lovely interpretation do not represent the charge density.

\section{Meaning of the elastic  Form Factor} My talks often include the statement that 
$G_E(Q^2)$  is NOT the  Fourier transform of charge density.  This is because the light masses of the quarks make necessary a 
 relativistic treatment. In that case, the  wave function is frame-dependent, initial and final states carry different total momentum and therefore are not the same. No square of a wave function enters, so there is  no density interpretation.

These difficullties can be handled using light front coordinates,   or equivalently working in the infinite   momentum frame.
The ``time" is $x^+=(ct+z)/\sqrt{2}=(x^0+x^3)/\sqrt{2}$. This can be thought of as the time in a frame moving with speed near $c$ in the negative $z$ direction. The canonically conjugate ``time" evolution operator $p^-=(p^0-p^3)/\sqrt{2}$. ``Space" $x^-=(x^0-x^3)/\sqrt{2}$ and momentum $p^+=(p^0+p^3)/\sqrt{2}$. The transverse coordinates position ${\bf b}$ and momentum ${\bf p}$ are as before. These coordinates are used to analyze form factors,  deep inelastic scattering and are the language of 
 GPDs and TMDs.

There is just one more thing to know about the light front. This is the kinematic subgroup of the  Poincar\'{e} group. Lorentz transformations defined by a velocity ${\bf v}$ in  the transverse direction do not involve interactions. In that case${\bf k}\to{\bf k}- 
{\bf v} k^+$. This is just like usual the Galilean transformation with $k^+$ playing the role of a  mass. If the  momentum transfer is in the perpendicular  direction, then the density is a two-dimensional  Fourier transform. This also requires 
taking $q^+$ of  the virtual photon momentum to vanish. Then
$-q^2=Q^2={\bf q}^2$. Using $q^+=0$ insures that the form factor involves transitions between Fock space components of the same number of partons. Thus one can measure a true density.

\section{phenomenology}
We used a model proton wave function that obeys Poincar\'{e} invariance. This wave function uses light front variables, so the boost caused by the absorption of a virtual photon in the transverse direction is included as a two-dimensional Fourier transform. A key feature is that Dirac spinors are used instead of Pauli spinors. Thus quarks carry orbital angular momentum. The 1995  model predicted the rapid decrease of the proton's $G_E/G_M$ and the related flat nature of 
$QF_2/F_1$.  See Figs. 10,11 of \cite{me1} and  Fig. 6 of \cite{me2}. The model consists  of three relativistic constituent quarks with a wave function
constructed so as to satisfy Poincar\'{e} invariance. It is essentially the non-relativistic quark model but the spinors are Dirac spinors, not Pauli  spinors. The  model lower components of Dirac spinors account for the non-zero
quark orbital   angular momentum. Given the presence of orbital angular momentum, I was asked about the shape of the proton, Does orbital angular momentum imply that the proton has  a non-spherical shape?
The Wigner-Eckart theorem tells us that the proton has 
no quadrupole moment. However, the non-spherical shape can be  revealed 
with spin dependent densities  SDD.
         
\section{spin-dependent density -non-relativistic example}
The example is in coordinate space.  Consider a non-relativistic proton, bound by a rotationally-invariant  potential to a
spherical core,  in a $p_{1/2}$ state. The  $J^\pi=1/2^-$ wave function can be written as 
\begin{eqnarray}
\langle{\bf r}_p|\psi_{1,1/2.2s}\rangle=R(r_p) 
{{\bfsigma}}\cdot \hat{\bf r}_p | s\rangle,
\end{eqnarray}
where $s$ is the $z$ projection of the total angular momentum and $|s\rangle$ is a Pauli spinor. The ordinary  density $\rho(r)$ is the matrix element of the density operator $\delta({\bf r}
-{\bf r}_p)$ in the $p_{1/2}$ state.  Thus the density
\begin{eqnarray}
\rho(r)=\int d^3r_p\langle\psi_{1,1/2.2s}|{\bf r}_r\rangle \delta({\bf r}
-{\bf r}_p)\langle{\bf r}_p|\psi_{1,1/2.2s}\rangle=R^2(r)
\end{eqnarray}
is spherically symmetric. It depends only on the distance from the origin. On the other hand, if one asks for the 
probability $\rho({\bf r},{\bf n})$ that the proton is at a position ${\bf r}$ and has a spin in a chosen direction $\bf n$, the spin-dependent-density, then 
\begin{eqnarray}
\rho({\bf r},{\bf n})&=&\int d^3r_p\langle\psi_{1,1/2.2s}|{\bf r}_p\rangle \delta({\bf r}
-{\bf r}_p){1\over2}(1+\bfsigma\cdot{\bf n})\langle{\bf r}_p|\psi_{1,1/2.2s}\rangle \nonumber\\
&=&{1\over2}R^2(r) \langle s|{{\bfsigma}}\cdot \hat{\bf r} (1+\bfsigma\cdot{\bf n})   {   {\bfsigma}}\cdot \hat{\bf r}|s\rangle
\end{eqnarray}
There are all kinds of interesting shapes. Define $\theta$ as the angle between the direction of polarization and the vector 
${\bf r}$. Then if ${\bf n}$ is parallel to the polarization direction the spin-dependent density $\sim\cos^2\theta$, while for 
the perpendicular case one gets $\sim sin^2\theta$

The
wave function of \cite{me1} was expressed in terms of  momentum space. So I computed the spin-dependent density in momentum space \cite{shape1,shape2}.  
 For quarks of high momentum, peanut and donut shapes are obtained for quark spins parallel and antiparallel to the direction of the nucleon polarization. This was an amusing outcome of a model, but there were obvious questions. Can one avoid the use of a model to discuss the proton shape. What is the relation between coordinate and momentum space densities? How can these shapes be  measured. Can one use the lattice and/or experiment? These questions are answered below.
 But we first need to discuss model independent versions of something simpler.
\section{Model Independent transverse charge densities}
Once can define an honest density $\rho_\infty(x^-,b)$  that is a matrix element of a true density operator \cite{mypion}. This is given by
\begin{eqnarray}\rho_\infty(x^-,b)=\langle p^+,{\bf R=0},\lambda|\sum_q e_qq^\dagger_+(x^-,b)q_+(x^-,b)|p^+,{\bf R=0},\lambda\rangle,
\end{eqnarray}
where the state $| p^+,{\bf R=0},\lambda\rangle$ is localized in transverse coordinate space and is a sum of eigenstates of ${\bf p}$, and $q_+$ is the independent part of the quark-field operator.  This equation along with the relation
$F_1=\langle p^+,{\bf p'},\lambda|J^+(0)| p^+,{\bf p},\lambda\rangle$ leads immediately to the transverse density:
\begin{eqnarray}
\rho(b)\equiv\int dx^-\rho_\infty(x^-,b)=\int {QdQ\over2\pi} F_1(Q^2)J_0(Qb)
\end{eqnarray}
As shown in Ref. \cite{myneutron} the transverse charge density is negative at the center, rises to positive values and then falls to negative values at large values of $b$.  For the neutron the negative definite nature of $F_1$ yields immediately the surprising result that 
$\rho(b=0)<0$. The central charge density of the neutron is negative. This seems to contradict previously held notions.

\section{Interpretation of the neutron transverse charge density}
Various questions arise concerning the origin of the negative central density. 
We used  \cite{ma} models of generalized parton distributions at zero skewness  to relate the behavior of deep inelastic scattering quark distributions, evaluated at high $x$, to the transverse charge density evaluated at small distances. We obtain an interpretation of the negative central  transverse charge density of the neutron. The $d$ quarks dominate the neutron structure function for large values of Bjorken $x$, where the large momentum of the struck quark has a significant impact on determining the center of momentum, and thus the ``center'' of the nucleon in the transverse position plane. 
The transverse charge densities discussed here involve $u-\bar{u}$ and $d-\bar{d}$. So a $\pi^-$ at the center
would decrease the former and increase the latter.

\subsection{Return of the Cloudy Bag Model}

We investigated \cite{rm1} the connection between the Breit and infinite momentum frames and show that when the nucleon matrix element of the time component of the electromagnetic current, which yields $G_E$ in the Breit frame, is boosted to the infinite momentum (or light-front) frame, the quantity $F_1$ is obtained. 

We also modeled  \cite{rm2} the physical nucleon  as a bare nucleon surrounded by a pion cloud to examine its implications for the neutron form factor, $F_1$, and the corresponding transverse charge density in the infinite momentum frame. Two versions, one with a tunable Pauli-Villars parameter, and one with light-front cloudy-bag pion-nucleon form factors, were examined. A qualitative agreement with the experimental form factor (negative definite for all $Q^2$) and transverse charge density are achieved when the nucleon is treated as having a finite extent. The bare nucleon must have a finite extent if this model is to account for a negative definite $F_1$, and consequently a negative transverse charge density of the neutron at its center.  In this model, the only negative charge resides on the virtual $\pi^-$, so  the negative charge density  at $b=0$  is caused by pions that penetrate to the center of the nucleon as in the CBM.

The combination of these two results indicates that, if one looks at the cloudy bag model in the infinite momentum frame, 
it has a negative central transverse density.

\section{Shapes of the proton through TMDs and Generalized Transverse Densities}

I've discussed how spin-dependent quark densities, matrix elements of specific density operators in proton states of definite spin-polarization, indicate that the nucleon may harbor an infinite variety of non-spherical shapes. We showed \cite{expshape}
 that these matrix elements are closely related to specific transverse momentum dependent parton distributions accessible in the angular dependence of the semi-inclusive processes electron plus proton goes to electron plus pion plus anything, and the Drell-Yan reaction proton plus proton goes to a lepton anti-lepton pair plus anything. New measurements or analyses would allow the direct exhibition of the non-spherical nature of the proton, in particular if the TMD $h_{1T}^\perp$ is measured to be non-zero. The TMDs are basically densities in  momentum space \cite {transverse}.

The coordinate-space spin-dependent density is obtained from  a  light-front version of the spin-dependent density operator:  $q^\dagger_+ (0,{\bf b}) {1\over
2} (1+{\bf n}\cdot\bfgamma)q_+(0, {\bf b}),$
where ${\bf n }$ is an arbitrary transverse direction \cite{(49)}. Evaluating the matrix element
of this quantity produces the spin-dependent density, which can be thought of as
the $x$-integrated version of the coordinate space results of \cite{shape1}. The proton is
non-spherical if the function $\tilde{A}''_{T10}$ of \cite{(50)} is non-vanishing, as recent lattice calculations
show \cite{(115)}.

\section{Summary}
Form factors,  GPDs, and TMDs, can now be analyzed and understood from a unified light-front formulation. In this talk, we reviewed how 
electromagnetic form factors of the neutron are used to show that the central transverse density is negative.
This may be  consistent with the ideas of the
 Cloudy Bag Model. Additionally, the non-vanishing of the coordinate-space spin-dependent density, as found by lattice calculations  shows that the 
proton is not round. 
Experiment can whether or not proton is round by searching for $h_{1T}^\perp$ in semi-inclusive electropion production reactions, and finding a non-zero value.




\begin{theacknowledgments}
 This work is partially supported by the USDOE.  I would also like to thank Tony Thomas for many useful, inspiring and provocative conversations over the years. I thank W. Melnitchouk for comments on the manuscript.
\end{theacknowledgments}




\bibliographystyle{aipproc}

\end{document}